\documentclass[12pt, a4paper]{article}

\usepackage{hyperref}
\usepackage{xcolor}
\usepackage{graphicx}
\usepackage{subcaption}
\usepackage[normalem]{ulem}

\usepackage[utf8]{inputenc}
\usepackage[a4paper,margin=2cm, marginparwidth=2cm]{geometry}

\providecommand{\href}[2]{\texttt{#2}}
\providecommand{\url}[1]{\texttt{#1}}

\providecommand{\eqref}[1]{(\ref{#1})}
\newcommand{\figref}[1]{Fig.\ \ref{#1}}
\newcommand{\secref}[1]{Section \ref{#1}}

\newcommand{\vdef}[1]{\textit{#1}}

\begin{document}

\begin{center}
{\Large\textbf{Understanding complexity via network theory: \\a gentle introduction}} \\[6pt]
 {\large {Vaiva Vasiliauskaite}}$^{1,2}$ and 
 {\large \href{https://www.imperial.ac.uk/people/f.rosas}{Fernando E. Rosas}}$^{1,3,4}$ \\[6pt]
\href{http://complexity.org.uk/}{Centre for Complexity Science}$^1$\\
\href{http://www3.imperial.ac.uk/theoreticalphysics}{Theoretical Physics Group, Department of Physics}$^2$\\
Center for Psychedelic Research, Department of Brain Science$^3$\\
Data Science Institute$^4$\\
\small{Imperial College London, London SW7 2AZ, U.K.}
\end{center}

\begin{abstract}
Network theory provides tools which are particularly appropriate for assessing the complex interdependencies that characterise our modern connected world. This article presents an introduction to network theory, in a way that doesn't require a strong mathematical background. We explore how network theory unveils commonalities in the interdependency profiles of various systems, ranging from biological, to social, and artistic domains. Our aim is to enable an intuitive understanding while conveying the fundamental principles and aims of complexity science. Additionally, various 
network-theoretic tools are discussed, and numerous references for more advanced materials are provided.
\end{abstract}

\renewcommand{\thefootnote}{\arabic{footnote}}

\section{Introduction}


\emph{Interdependency}, represented by links that connect different elements, is a fundamental notion that is widely applicable when studying complex systems.
For example, a group of friends are connected via friendship links, whereas cities are linked via train lines. Similarly, an entity---a city, a person, a cell---can be linked to other entities in a wide variety of ways: physical vicinity, friendship, common genes, shared ideology, and so on. Interestingly, underneath the diversity of entities and variety of their linkages, there tend to exist interesting commonalities in the organisational structure. 
In fact, these complex, seemingly random systems, as dissimilar as biological, economic, sociological and physical systems, seem to share structural regularities. It is those regularities what \emph{network theory} attempts to capture.

Network theory also provides efficient tools to exploit the so-called ``big data". While communication technologies enable the gathering of unprecedented (and ever-increasing) amounts of information, data by themselves don't tend to reveal meaningful insights easily. 
Why there is so much inequality in the distribution of wealth in many societies? What makes some diseases spread much faster than others? How important is a particular species for maintaining the stability of an ecosystem? These questions are challenging as they involve not only properties of the object itself, but also its relationships with the surroundings. Addressing such questions is where network theory methods, combined with big data, shine at their fullest.

Because network theory is built upon analytical foundations, some mathematical knowledge is often needed to access many of the available resources. To help fill this gap, this article provides a gentle introduction to network theory without mathematical formulas. We give an intuitive presentation which avoids unnecessary technicalities, yet still conveys the spirit and fundamental principles of complexity science through the network viewpoint. Hence, it is our hope that this overview provides the reader with conceptual and suggestions for practical applications, and will enable them to employ this framework in order to make sense of our complicated interconnected world.

The rest of this article is organised as follows. First, \secref{scomplexity} introduces the fundamental notions of complexity science, and sets the conceptual basis for the rest of the review. Then \secref{snetworks} explores the fundamental principles of network theory, and discusses a set of examples that illustrate these principles. \secref{snetworkfeatures} presents tools to characterise different networks and better understand their various properties. \secref{snetwork_architecture} reviews a number of network architectures, which are used for network modelling. Finally, \secref{sdiscussion} concludes this article with future perspectives of network theory, gives suggestions of further reading, and also reviews a number of limitations of this approach.

\section{Complexity}\label{scomplexity}

In order to prepare the ground for subsequent sections, here we introduce some fundamental concepts from complexity science. In the following we discuss the difference between complex and complicated, the relationship between reductionism and complexity, and the role of synergy and emergence. Please note that some of our terminology is not standard.

\subsection{Complicated does not implies complex}

Complex and complicated might sometimes be used interchangeably in our daily life; however, it is useful to draw a technical difference between these two terms. 

We define a system to be \vdef{complicated} if it cannot be characterised by any short description. As descriptions might depend on the language employed, it is important to distinguish between apparent and intrinsic complicated-ness: the former takes place when a long description is a consequence of an ill-suited language, while the latter corresponds to the case when there is no language that allows a short description. A classic example of apparent complexity is describing a square shaped wave: using Fourier analysis requires an infinite number of sine-waves to describe a square wave, while an alternative description based on modern wavelet analysis is much briefer. In contrast, intrinsically complicated objects cannot be properly described succinctly in any language. 

A simple approach to study intrinsic complicated-ness is based on the notion of ``inability to compress,'' this being related to high Shannon entropy, c.f.~\cite{cover2012elements}. However, this approach does not work directly on concrete objects but over models that may generate such objects. Unfortunately, inferring models from objects always requires strong assumptions, and hence any study based on Shannon's entropy depends critically on them. A more direct approach to study complicated-ness that avoids such assumptions relies on the idea of \textit{Kolmogorov complexity}, which quantifies the length of the shortest computer program capable of generating the object of interest~\cite{li2008introduction}. Unfortunately, the Kolmogorov complexity is usually uncomputable in most practical scenarios, and hence one needs to rely in approximations~\cite{li2008introduction}.

\subsection{Complexity science as a complement of reductionism}

A big portion of contemporary science is driven by the spirit of reductionism, which under the motto of \textit{divide et impera} studies hard problems by dividing them into easier problems. When studying a multi-agent system, reductionist approaches first focus on understanding the nature and properties of each agent in isolation, and only when this is solved, it proceeds to see how agents act together. 

While reductionism has provided satisfactory results in a wide range of scenarios, it has proven ineffective in a number of important problems. For example, claims exist that a number of important social and economical issues of our modern world---including climate change and wealth inequality---are unlikely to be solved by reductionist approaches~\cite{senge2010necessary}. As another prominent example, some neuroscientists argue that high brain functions and consciousness might be related to irreducible relationships that take place in the orchestrated activity of a number of different brain areas, a result that would be invisible to reductionistic approaches~\cite{kelso1995dynamic,chalmers1996conscious}.

\begin{figure}
    \centering
    \includegraphics[width = 14cm]{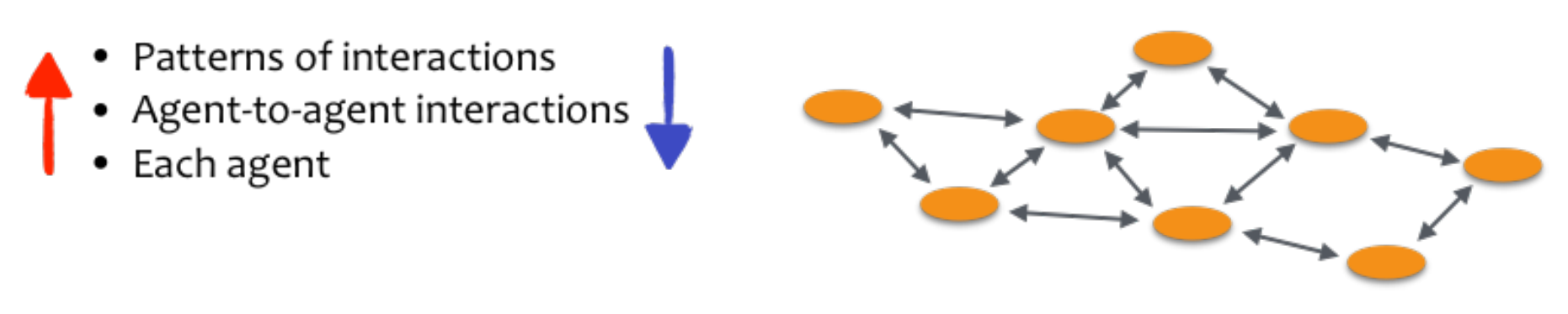}
    \caption{A network might represent an abstract multi-agent system, and the red and blue arrows illustrate reductionism and emergentism. Reductionism corresponds to the red arrow, which builds up from individual agents; emergentism is represented by the blue arrow, which starts considering patterns of collective interactions and only afterwards attempts to understand specifics aspects of the system.}
    \label{f1}
\end{figure}

Complexity science provides a view that complements reductionism (see~\figref{f1}). In effect, while reductionism studies first the parts and then their interactions, complexity science first considers the patterns of interactions to only then try to understand individual agents. Accordingly, we define a \vdef{complex system} as one that cannot be fully understood via reductionist analyses, being better suited to analyses that take as a start point the patterns of interactions.

\subsection{Synergy and networks}
\label{sec:synergy}

Technically, \emph{synergistic phenomena} are those than can be seen in the whole of a system but not in the parts~\cite{rosas2016understanding,Rosas2019}. As a matter of fact, it is the presence of synergistic phenomena the reason why reductionist approaches sometimes fail. Therefore, synergy lies at the very heart of complex systems.

Some synergistic phenomena are especially persistent, even gaining apparent agency. Those cases are referred to as \emph{emergent}, which alludes to the fact that macroscopic variables with apparent agency ``emerge'' spontaneously from the underlying substrate. The literature about emergent phenomena is diverse and unclear, and the subject is more discussed in the literature of philosophy than in quantitative sciences. For good introductions to the theory of emergent phenomena please refer to~\cite{holland2000emergence}. Recent attempts to quantify emergence include the work reported in~\cite{seth2010measuring,hoel2013quantifying,mediano2019beyond}.

A popular method to study collective properties associated with synergistic phenomena is network theory.  Networks allow us to focus \textit{less in the matter and more in the patterns}, i.e.\ in the ``shape'' of the arrangement of interdependencies across a system of interest. How this is done, and it implications, is the subject of the rest of this text.


\section{Networks}\label{snetworks}

This section covers the fundamental concepts of network theory. After a brief introductory discussion outlined in \secref{sec:31}, in \secref{sbasicdef} basic definitions 
of complex network analysis are introduced. In \secref{sec:33} we present a selection of real-world networks, which in many ways motivate and illustrate the main concepts introduced in previous subsections. 

\subsection{What is a network?}\label{sec:31}

A network is often not an object in and of itself, but a way to represent a system. In effect, we believe the question \textit{``is this a network?''} is normally ill-posed, whereas \textit{``would it make sense to model this as a network?''} would be a more useful one. In effect, in this presentation we consider networks to be a tool or a metaphor, i.e.\ a way of describing systems composed of many sub-units. 
Hence, a network is an abstraction that depicts the interdependencies that exist in a given system of interest. 

Just as a map neglects aspects of the territory in order to focus on particular features of interest, a network takes away most of the richness of individual sub-units of the system and only retains the structure of their interdependencies. Networks allow us to study the collective properties of these dependencies. Additionally, as seemingly unrelated systems might have similar networks of interdependencies, networks sometimes allow us to establish non-trivial relationships between systems that share similar dependency structures.

\subsection{Basic definitions}\label{sbasicdef}

A \vdef{network} is a collection of vertices and edges: a \vdef{vertex} (or a \vdef{node}) represents a particular sub-unit of the system, and \vdef{edges} connect sub-units/vertices that interact with each other. Networks, hence, depict \textit{bilateral relationships} between objects. Note that in general each vertex has a unique identity, but no further internal structure.

Many analyses start from a dataset, which in itself has no network representation. Therefore, a crucial initial task for a researcher is to define nodes and a relationship criterion to establish edges. The goal is to find a representation which beholds the information needed to address a particular question of interest. The reduction of data that takes place when building a network representation leads to an inevitable loss of information. This is a price one needs to pay in order to obtain a simplified representation of the interdependencies seen in the data.

Networks are studied in mathematics under the name of \vdef{graphs}. In effect, \textit{graph theory} is a branch of pure mathematics that is concerned with general properties of graphs. Network science builds on top of graph theory, greatly expanding its scope by including methods and approaches from statistical physics.

The type of a network described so far is the most basic: there is only one type of node, and the connections between nodes are binary: an edge either exists or does not. However, in some situations one might want to employ more sophisticated models which provide more flexibility. 
For example, one can assign values to the network's edges and obtain a \vdef{weighted network}. The weight of an edge represents its strength: the larger the weight, the stronger the connection between nodes. Moreover, one can allocate a sign to these values, and generate a \vdef{signed network} whose edges can have either positive or negative weights. Similarly, assigning directionality to edges makes a network \vdef{directed}. Furthermore, if one wants to describe non-binary relations between nodes, \vdef{hypergraphs} can be employed. In such network an edge joins more than two nodes together. Lastly, \vdef{multidimensional networks} entail several different types of connections. 

Labels and categories can also be added to nodes. For example, \vdef{bipartite networks} consist of exactly two types of nodes, connected in such a way that edges can only take place between nodes of different types. \vdef{Multilayer networks} are a generalisation of bipartite networks in which multiple types of nodes exist. Finally, one can also assign a location to the nodes (and edges) and obtain a \vdef{spatial network}.

The definition of a network is very flexible. Thus this approach enables one to model a wide range of real systems. Given a complex system one simply needs to find the most natural network representation. 

\subsection{Examples of networks}\label{sec:33}


There are numerous examples of complex systems that can be modelled as networks. Network analysis has been useful in studies of technological, social, biological, chemical, physical structures and systems. 

Social scientists have used the concept of a social network since the early $20^\textrm{th}$ century. In effect, many network analysis tools and methodologies root back to and build upon sociometry (or social network analysis, SNA), a term coined by social scientists for a study of complex social systems~\cite{M34,F04}. A social network is typically a network composed of individuals (or groups of individuals) as nodes and their interactions, represented as edges~\cite{OR02}. Perhaps the simplest social network is composed of individuals, represented as vertices, joined by edges if they are friends---a so-called ``friendship'' network. However friendships are not the only type of social relations between individuals. One can similarly model collaborative relationships between artists~\cite{GD03} or scientists~\cite{P10,GUSA05,M04}, or even evaluate how much our genes impact our social interactions~\cite{FDC09}. In an organisation, people usually have a position in an official hierarchy at work, so a worker may be linked to his/hers boss, or linked to people that sit nearby in the office. It has been shown that important links could come from occasional meetings of people in other parts of the organisation or even in rival companies (the so-called ``strength of weak ties"~\cite{G73}).

Networks are also useful for characterising the interdependencies that take place in biological systems. 
The structures we observe in nature unfold layers upon layers of complexity, ranging from microscales that explain cellular behaviour, to macroscales where they give insights about animal interactions in an ecosystem~\cite{cofre2019comparison}. Given high-throughput experimental results, nowadays scientists are able to model inter- and intra-cellular interactions, relations of genomes, and observe biomass flow from one species to another as food webs. For instance, protein-protein interactions have been extensively studied as networks~\cite{CFMV05,NYP12}. In this network, proteins are represented as nodes, which are pairwise connected if they participate in the same interaction. 

\begin{figure}
    \centering
    \includegraphics[width = \linewidth]{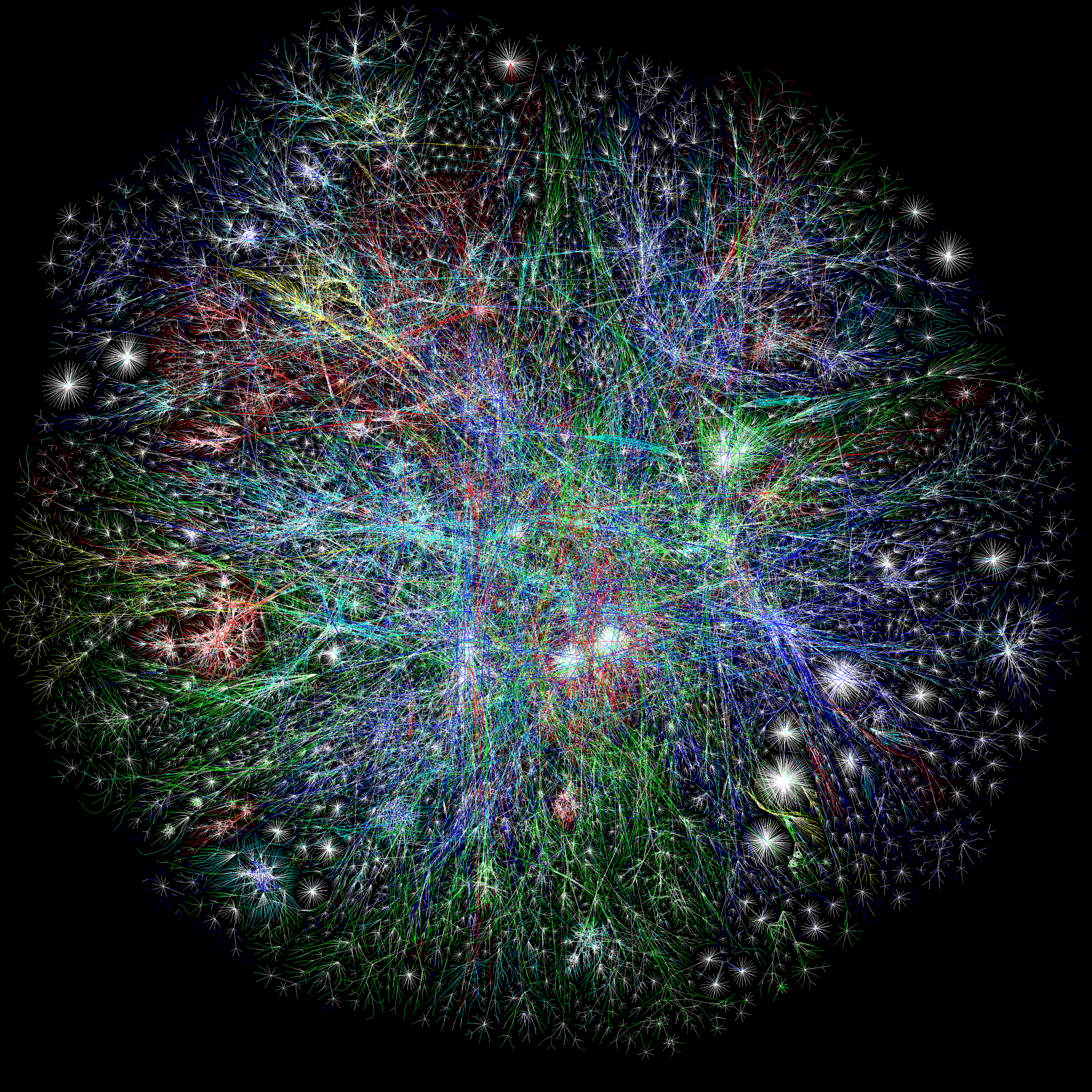}
    \caption{Partial map of the Internet based on data from November $24^{\mathrm{rd}}$ 2003, taken from \href{http://opte.org/maps}{http://opte.org/maps} (accessed on February $5^{\mathrm{th}}$ 2020). Each line is drawn between two nodes, representing two IP addresses. The length of the line is indicative of the delay between the two connected nodes. Lines are colour-coded based on Class A allocation of IP space (Asia Pacific - Red
Europe/Middle East/Central Asia/Africa - Green
North America - Blue
Latin American and Caribbean - Yellow
RFC1918 IP Addresses - Cyan
Unknown - White).}
    \label{finternet}
\end{figure}

Let us move from ``natural'', organic networks, to cyber-networks, which are numerous as well. A remarkable example of a digital network is the Internet (see \figref{finternet} for visualisation). In this network, nodes are physical computers, linked by cables. The large-scale topology of the internet is complex, as it is a self-organizing system whose properties cannot be traced back to any global blueprint or chart~\cite[Chapter 3]{PV04}. The World Wide Web is a virtual network of information, living on the structure of the Internet. One can find many different types of networks in this system. Perhaps the most obvious one is the network of web pages as vertices, in which an edge indicates a hyperlink from one web page to another. Alternatively, one can look at a ``semantic'' network of the World Wide Web in which two web pages are connected with an edge if their most common words are similar. Lastly, one can link two pages with an edge if they both point to the same third web page, to produce a ``co-citation'' network. 

One may also construct numerous networks by considering spatial positions of some elements---spatial networks, described briefly in \ref{sbasicdef}. A so-called ``urban'' network can be constructed by considering street intersections as network nodes. One obvious network of such nodes can be obtained by calling portions of streets between intersections to be edges. A bit less-intuitive is a street network, in which the spatially straight street sections are vertices and intersections between them are edges.

These are just a few examples of the numerous systems that can be studied as networks.

\section{Features, properties and tools for networks}\label{snetworkfeatures}

A network model enables mathematical and numerical analyses which give insights into the observed structures. One can use them to understand questions related to the origin of the network (what processes lead to the observed network structure?), evolution of the network (i.e.\ its growth in time or space), dynamical processes on the network (e.g.\ epidemics). The tools presented in the following discussion may also help characterise different types of networks and allow us to study their emergent properties.

Results of a given analysis strongly depend on the scale at which the network is studied. If one looks at a network at the level of individual node relationships, macroscopic, local, and somewhat static properties about the network are revealed. 
On the other hand, looking at higher-order structures one gains information about non-local interaction of nodes, which can further be linked to some macroscopic dynamics. So it may be possible to demonstrate how the microscopic (bilateral) interactions lead to cooperative phenomena and emergent properties of some dynamical processes, such as diffusion, broadcasting, synchronisation, if present in the system.

We finish this section by discussing how, having learnt about the structure of the observed networks, synthetic networks that mimic some certain important features can be built.

\subsection{Visual inspection}

A simple first step that often helps towards understanding a particular network is direct visual inspection. Besides being a sanity-check of the data, visual inspection can provide some basic insights about the network of interest. In effect, a well-done presentation of any data can give useful insights, or at the very least be visually pleasing. Please note that network visualisation is both an art as well as a science; automated methods alone rarely produce the most effective visualisations.

Network visualisation software are numerous, much of which are also freely available online. Some examples of stand-alone software for network visualisation inlude \href{https://gephi.org}{Gephi}, \href{
https://www.graphviz.org}{Graphviz}, \href{https://www.neo4j.com}{Neo4j}, \href{https://www.visone.info}{Visone}, \href{https://www.smrfoundation.org/nodexl/}{nodexl}, \href{https://cytoscape.org/}{cytoscape}. Various network-oriented libraries, available in different programming languages also include visualisation functionality. Examples of such libraries are \href{https://www.boost.org/}{Boost} (C++), \href{https://igraph.org/}{igraph} (Python and R), \href{https://networkx.github.io/}{networkx} (Python), \href{http://jung.sourceforge.net}{JUNG} (Java).

Representing a network graphically is a more complicated task than it may seem first, as vertices and edges do not in general have any coordinates in space. When drawing them one may place vertices anywhere, provided that one connects vertices faithfully. However, one might quickly find crossing edges, and eventually the visualisation can easily become confusing or even misleading. Large network visualisations, which sometimes don't convey much information, are sometimes called ``spaghetti monsters'' (the reader can decide if \figref{finternet} falls into this category or not).  These monsters can still be analysed using other techniques, which are discussed below.

\subsection{Differentiation}

The network structure describes how nodes are related directly through edges and indirectly through chains of relationships. So a natural question to ask is if nodes group together in differentiated clusters/communities. The ``grouping'' of nodes can be assessed in different ways, some of which are explored in what follows.

Here we present several network observables that allow us to differentiate network structures. On macroscales, one can compare the amount of clustering of nodes into groups via the modularity; whereas on microscale the information about the density of triangles as well as motifs reveals the characteristic local structure of a network. From these complementary angles, these approaches provide some light over the network's functional abilities.

\subsubsection{Clustering coefficient and motif analysis}

Measuring a \vdef{clustering coefficient} is a common first step towards assessing the structure of the links within a network. This quantity captures the propensity of nodes that are connected to a common node to be connected between themselves as well. For instance, in a friendship network, if ``a friend of a friend is also a friend", it is represented by a triangle. Therefore, 
the clustering coefficient operationalises this notion by counting for ``triangles'' (triples of nodes, all connected pairwise with edges). Technically, the clustering coefficient is defined as the ratio of actual triangles in a network and the number of potential triangles---three nodes connected by two or three edges. Social networks tend to have quite high clustering coefficients~\cite{N10}. The values of the clustering coefficient can range from zero to one, but for many (although not all) real world networks, the observed clustering coefficient is significantly larger than one would expect by chance. For instance, a co-authorship network of physics researchers has a clustering coefficient of 0.45. This value indicates that there is a $45\%$ probability that two neighbours of a vertex are neighbours themselves. If edges were placed at random in the network, the clustering coefficient would have yielded a mere 0.0023~\cite{N10}. Other networks that show larger-than-expected clustering coefficients include a film actor collaboration network~\cite{N06} and a network of company directors~\cite{DYB03}. Some networks, that do not seem to have a significantly clustered structure include a network software package dependencies, and metabolic networks~\cite[Table 8.1]{N10}.

Triangles, used in the aforementioned clustering coefficient, are not the only small groups of vertices that are commonly encountered in networks. Distinct, dominant (and usually small) structures of any form are collectively called \vdef{motifs}. A motif can be any set of nodes, connected in a specific way, given that we observe this arrangement in our network repetitively. 
The over-representation of specific motifs in networks, such as gene regulatory networks and neural networks, may signal their importance for the organisms involved~\cite{M04}. In particular, they may act as ``circuit elements", such as filters, important to perform certain tasks~\cite{N10}.

To study motifs, one first defines the motifs of interest and counts their appearance in a network of interest. To assess if these motifs are appearing significantly often, one counts their occurrence on a randomised network which acts as a null model\footnote{In randomised networks one does not expect to see any preponderance of particular patterns; see Section~\ref{snetwork_architecture} for a discussion on random graph models.}. One can declare the motif to be present in the network if the number of occurrences is much higher than in the randomised networks. 

Motifs have recently gathered much attention as a useful concept to uncover structural design principles of complex networks. 
Indeed, very frequently occurring motifs in networks such as genetic regulatory network are thought to be related to its functional properties and perhaps they are an the result of evolution, beneficial for the organisms involved~\cite{MSIKCA02}. More specifically, much experimental work has been devoted to understanding network motifs in gene regulatory networks. These networks control which genes are expressed in a cell in response to some biological signals.

\subsubsection{Modularity and community detection}\label{scommunitydetect}

If the clustering coefficient of a network varies significantly from one region to another in the network, it is likely that the network is \vdef{modular}. The characteristic feature of such networks is the presence of densely connected groups of vertices, known as communities or clusters, with sparser connections between these groups. Many real networks exhibit modular structures, including social networks, computer networks, and metabolic and regulatory networks~\cite{N06}.

To quantify how well defined these groups are, one may calculate the value of a so-called \vdef{modularity} function. The idea behind this modularity measure is that true community structure in a network corresponds to a statistically surprising arrangement of edges. So in modularity the number of edges falling within groups is compared to the expected number in an equivalent network with edges placed at random~\cite{N06} (other null models can be employed; we will discuss some of them in \secref{snetwork_architecture}). The more ``surprising" the arrangement of edges, the larger the modularity score. \vdef{Community} is intuitively defined as a group of nodes which are more closely connected with each other, rather than with the rest of the network. 

\begin{figure}
    \centering
    \includegraphics[width=0.5\linewidth]{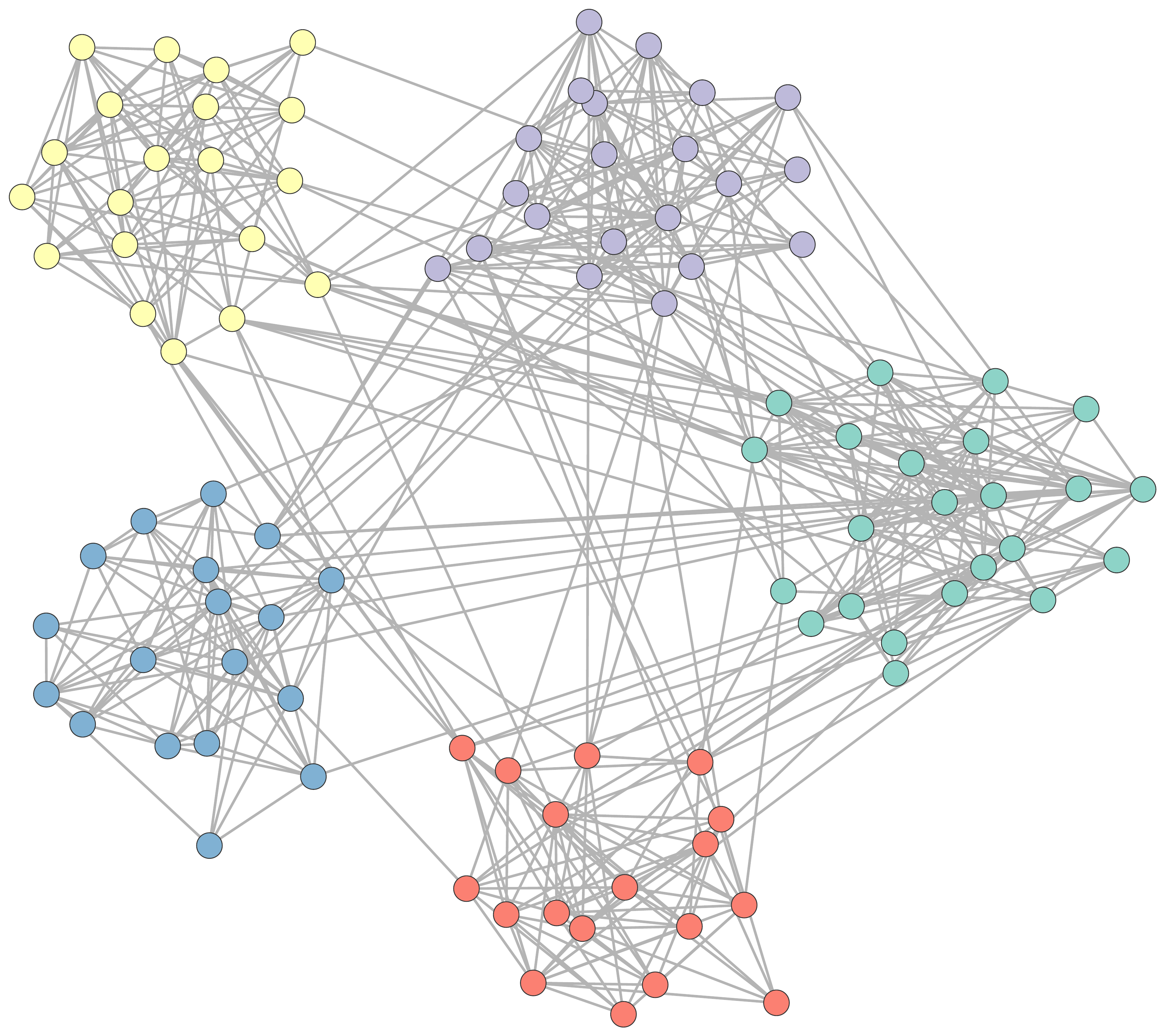}
    \caption{An example of a random modular network, composed of 100 nodes which have an average degree of 10.3. Degree distribution is Poissonian. Nodes in the same community are of the same colour. The network was generated using the methodology in~\cite{SSCB14}, and communities were found using \href{https://python-louvain.readthedocs.io/en/latest/}{Louvain algorithm}~\cite{BGLL08}). Modularity of this partition is $0.63$.}
    \label{example_community}
\end{figure}

Social networks are typical examples of graphs with communities. The word ``community" itself refers to a social context. People naturally tend to form groups, within their work environment, family, friends. We show an example of a network with apparent communities in \figref{example_community}. Biological networks also exhibit modular structures. For instance, protein interaction networks, fundamental for multiple processes in any cell, are also composed of parts that are more interconnected and less so. Communities are thought to correspond to functional groups, for instance, proteins having the same or similar functions, which are expected to be involved in the same processes~\cite{F09}. Some other examples of modular networks are: the World Wide Web, scientific collaborator networks, mobile phone communication networks~\cite{BGLL08}, and networks from social media websites such as Twitter or Facebook, see Section 17 in~\cite{F09}.

The ability to discover groups or clusters can be a useful tool to reveal structure and organisation within networks at a scale larger than that of a single vertex, or a triangle, or a small motif. If a network is small, a visual inspection is enough to find if communities are in the network and if present, what distinguishes the different groups. When the networks are large, algorithms are more practical to find where the communities lie. 

There are many different approaches to community detection. A classic, hard algorithmic problem, called graph partitioning, is a close predecessor of community detection algorithms. It is assumed that the number of underlying communities and so their typical size is known. The goal of graph partitioning is to find a partition of nodes such that the edges between those communities are minimal. Finding the exact solution is essentially impossible (it is in the category of \vdef{NP-hard} problems), so typically solutions to the graph partitioning problem are derived using heuristics or approximate algorithms. Graph partitioning is a fundamental issue in parallel computing, circuit partitioning and layout, and in the design of many serial algorithms, including techniques
to solve partial differential equations and sparse linear
systems of equations. For current state-of-the-art and the review of traditional graph partitioning methods please refer to~\cite{BMSSS13}.

When the number and size of communities are not known, the task we have at hands is known as community detection. An early attempt at identifying communities in a network was made by Girvan and Newman~\cite{GN02}, who developed their homonymous algorithm based on edge betweenness centrality for uncovering a hierarchical community structure in networks. Their algorithm is based on the maximisation of a modularity function, which compares the network to a null model. If the found communities are ``high-quality", the modularity function returns a high score. To find ``the best" communities, optimisation algorithms are used, such as Louvain~\cite{BGLL08}, Infomap~\cite{RAB08}, and so on.

One of the criticisms that community detection methods get is that some optimisation algorithms are subject to a resolution limit~\cite{FB06}: sometimes communities might remain undetected, whereas other times nodes might be grouped together even if they should not belong to the same community. Alternatively, methods based on machine learning are useful when the number of communities is known~\cite{SZ16}. To address the number of communities, recent work of Newman and Reinert~\cite{NR16} have proposed methods based on inference to estimate the number of communities in a network.

\subsection{Integration}

A concept that is complementary to differentiation is integration, which refers to the ``well-connectedtedness'' and ``navigatability'' of the network. Similar to the differentiation of a network, the intuitive notion of integration can be implemented in a number of different ways. We now examine some of them.

\subsubsection{Shortest paths}

A \vdef{walk} in a graph is a sequence of vertices, such that there is an edge from each vertex to the next vertex in the sequence. Intuitively, one can imagine a walk as the ``trail of a walker'' on the graph. A walker is allowed to make a step from one node to the next only if there is an edge connecting the nodes. 
A \vdef{path} is a walker's trajectory where she/he doesn't visit any node twice. One can characterise a given network in different ways according to various types of paths. For example, one can find the paths which are the shortest in length between given nodes.

The properties and statistics of various paths yield interesting insights on particular networks. For instance, the distribution of the shortest paths between nodes relates to the efficiency of communication in our network. More so, the longest of all shortest paths, the \vdef{diameter} of a network, represents a linear size of a network. Small diameter is also thought to be related to large \vdef{heterogeneity} of degrees and small amounts of synchronisation in the network~\cite{NMLH03}. Furthermore, as we will see in Section~\ref{snetwork_architecture}, so-called ``small-world" networks are defined as networks with large clustering coefficient, yet small diameter.

To study dynamics of processes on networks one can employ various stochastic methods, for example random walk techniques. Random walks are often used as a model for diffusion, and there has been intense research on the impact of network architecture on the dynamics of random walks. The finiteness of a network---along with properties such as degree heterogeneity, community structure, amongst others---can make diffusion on networks both quantitatively and even qualitatively different from diffusion on regular or infinite lattices. Thus the statistics and dynamical properties of a random walk statistics can reveal the structural properties of a network to us. Furthermore, different variations of random walks can be used to describe multiple dynamical processes on networks, such as diffusion, broadcasting, to name a few~\cite{MPL17}.

\subsubsection{Centrality}

In the simplest examples, nodes in a network are equivalent; they have no internal structure and represent the same type of entities.
However, the structural positions of nodes within a given network make them far from equivalent: some are positioned in structurally important (central) parts of a network, and some are tucked away in peripheral areas. The \vdef{centrality} of a node evaluates its positional importance with respect to the rest of network's topology. Centrality is thought to answer the question of which nodes are the more important ones---according to an agreed definition of importance.

Interestingly, the notion of centrality can be defined in multiple ways. To evaluate centrality of a node, various types of information about its position in a network and relation to other nodes are taken into account. 

The simplest centrality measure is degree: how many links are adjacent to a node. For example, Garfield used degree centrality in a citation network to quantify the quality of a publication by the number of citations it had~\cite{G73}. A node's betweenness centrality~\cite{F78} is defined as the ratio of shortest paths in the network that travel via that specific node. Closeness centrality is a measure of how long it takes to reach all the other nodes in the network from a specific node. It is defined as an inverse of a mean shortest path distance from that node to others. Other centrality measures are based on various types of walks. For instance, Katz centrality accounts for all paths, adjacent to a node and the score is proportional to the number of walks which end at the node from any starting vertex, scoring shorter walks higher than longer paths~\cite{K53}.

Centrality has found many uses in real-world networks. For example, Brin and Page developed PageRank centrality measure to rank websites based on their importance in the hyperlink network of the World Wide Web~\cite{BP98}. In this network, an edge between two websites represents that one website features a hyperlink to another. The more often the website is pointed to by important websites, the more central it is. This centrality measure was the fundamental building block of Google search queries. 

Bavelas~\cite{B48} first remarked in his experiments that centrality is related to group efficiency in problem solving. This idea of centrality being related to other group processes was well investigated in the following decades by sociologists~\cite{M66,P65,B65}, later adapted to other sciences and is used to study, for example, biological systems~\cite{A18} and is applied to recommendation systems~\cite{PBMW99}.

For further reading about centrality, refer to~\cite{LFH10} and references therein, as well as any general books, such as~\cite{N10}.

\section{Network Models}\label{snetwork_architecture}

The previous sections have given us plenty of tools to study networks of interest. 
Equipped with this knowledge, we now explore some popular models used in network theory, which have proven useful when studying diverse real-world scenarios. These models also allow us to deepen our understanding on how various network properties may affect the behaviour of a system.

Technically speaking, most network models are \vdef{random graphs}, in which some parameters are fixed but other features vary (e.g.\ by shuffling connections between nodes). One can think of a universe of equivalent networks---an ensemble---which determines each model; from this set one may pick a particular network and study its properties. Properties that one might fix include number of nodes, number of edges, degree of each node, diameter, clustering coefficient, etc.

In the following sections we review some of the most important network models in the literature of network theory.

\begin{figure}[!h]
\centering
\begin{subfigure}{.45\textwidth}
  \centering
  \includegraphics[width=.9\linewidth]{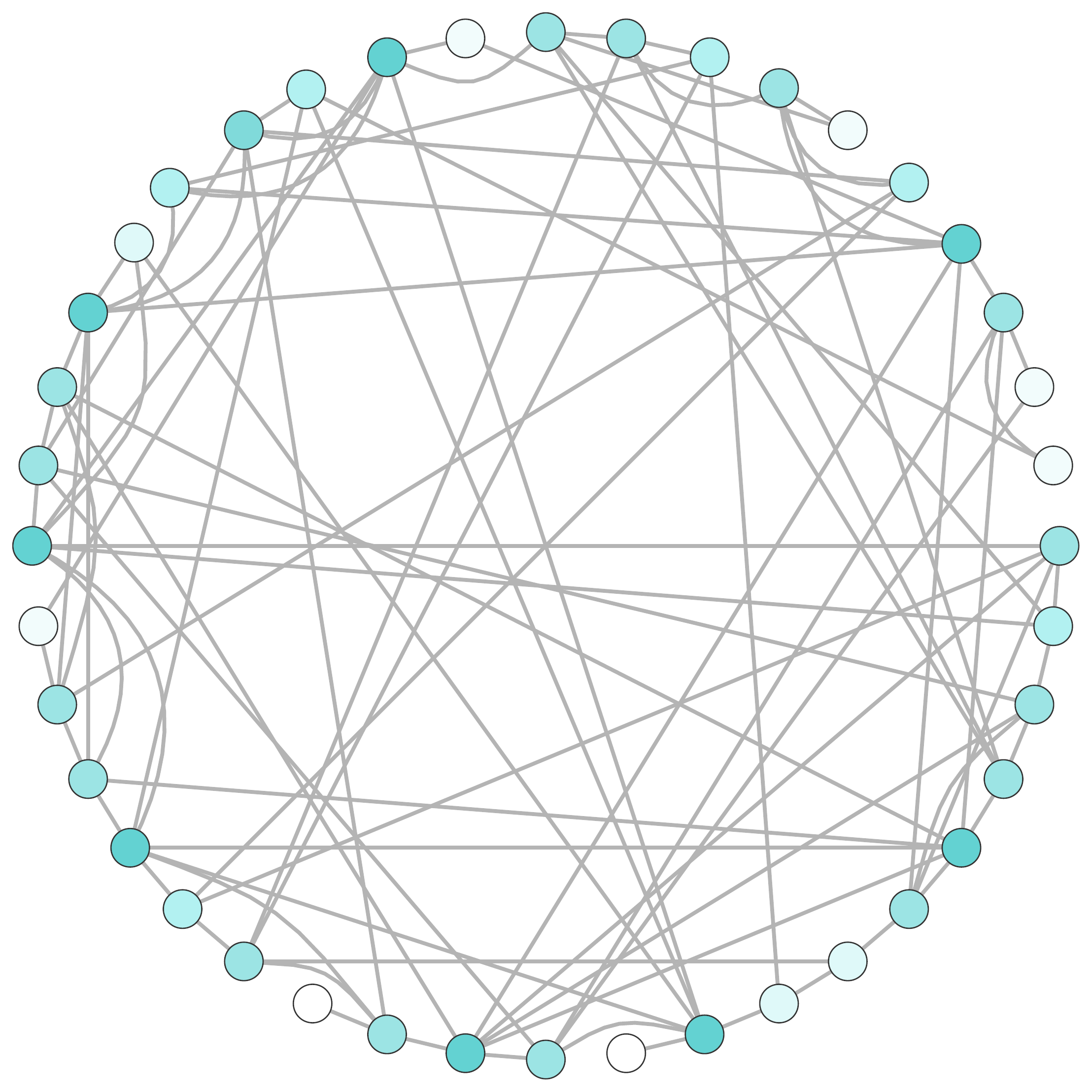}
  \caption{An Erd\"os-R\'enyi graph, obtained by connecting each pair of nodes with a probability of $13\%$.}
  \label{er}
\end{subfigure}%
\hspace*{.05\textwidth}
\begin{subfigure}{.45\textwidth}
  \centering
  \includegraphics[width=.9\linewidth]{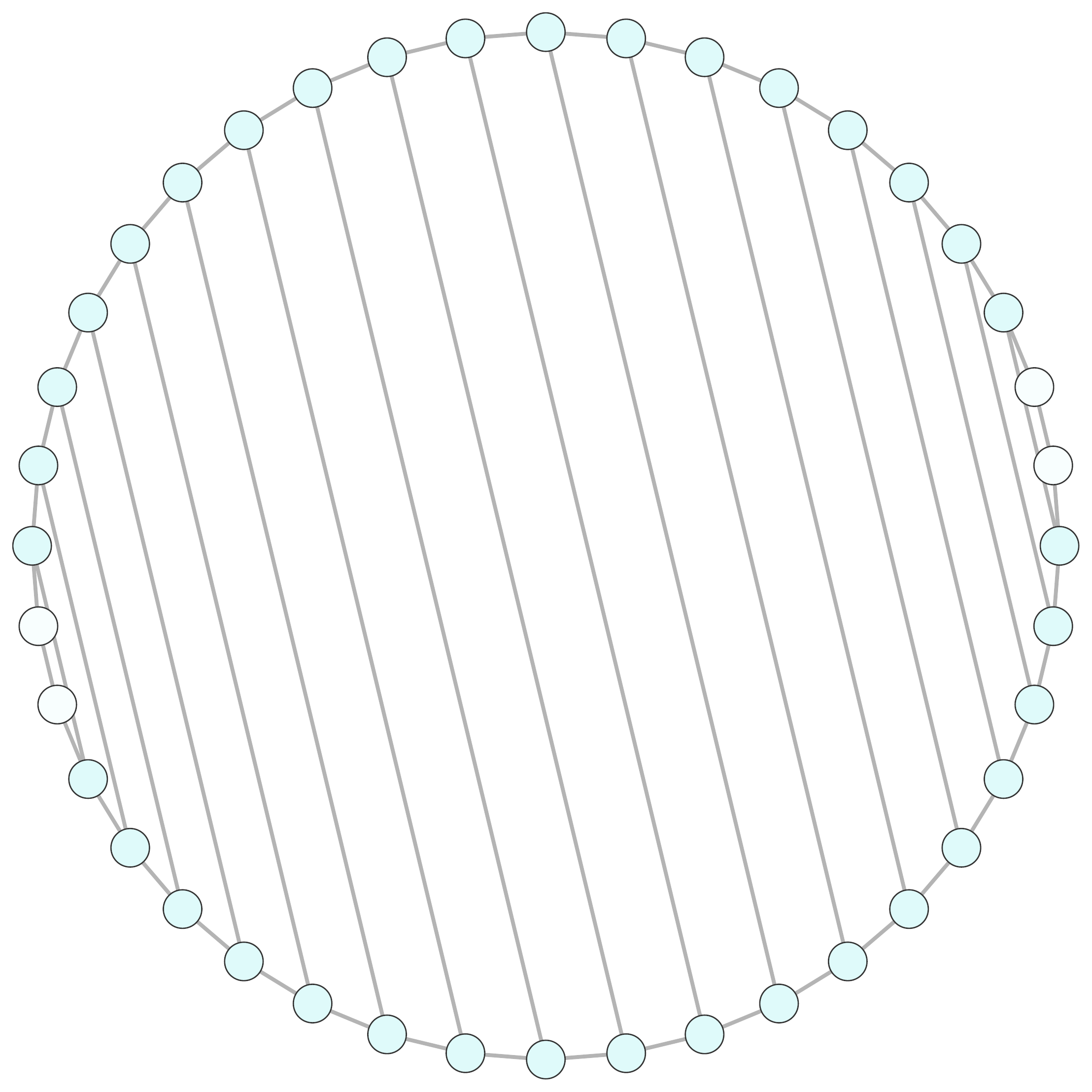}
  \caption{A 2-dimensional lattice graph, in which each node is connected to its three nearest neighbours.}
  \label{grid}
\end{subfigure}%

\begin{subfigure}{.45\textwidth}
  \centering
  \includegraphics[width=.9\linewidth]{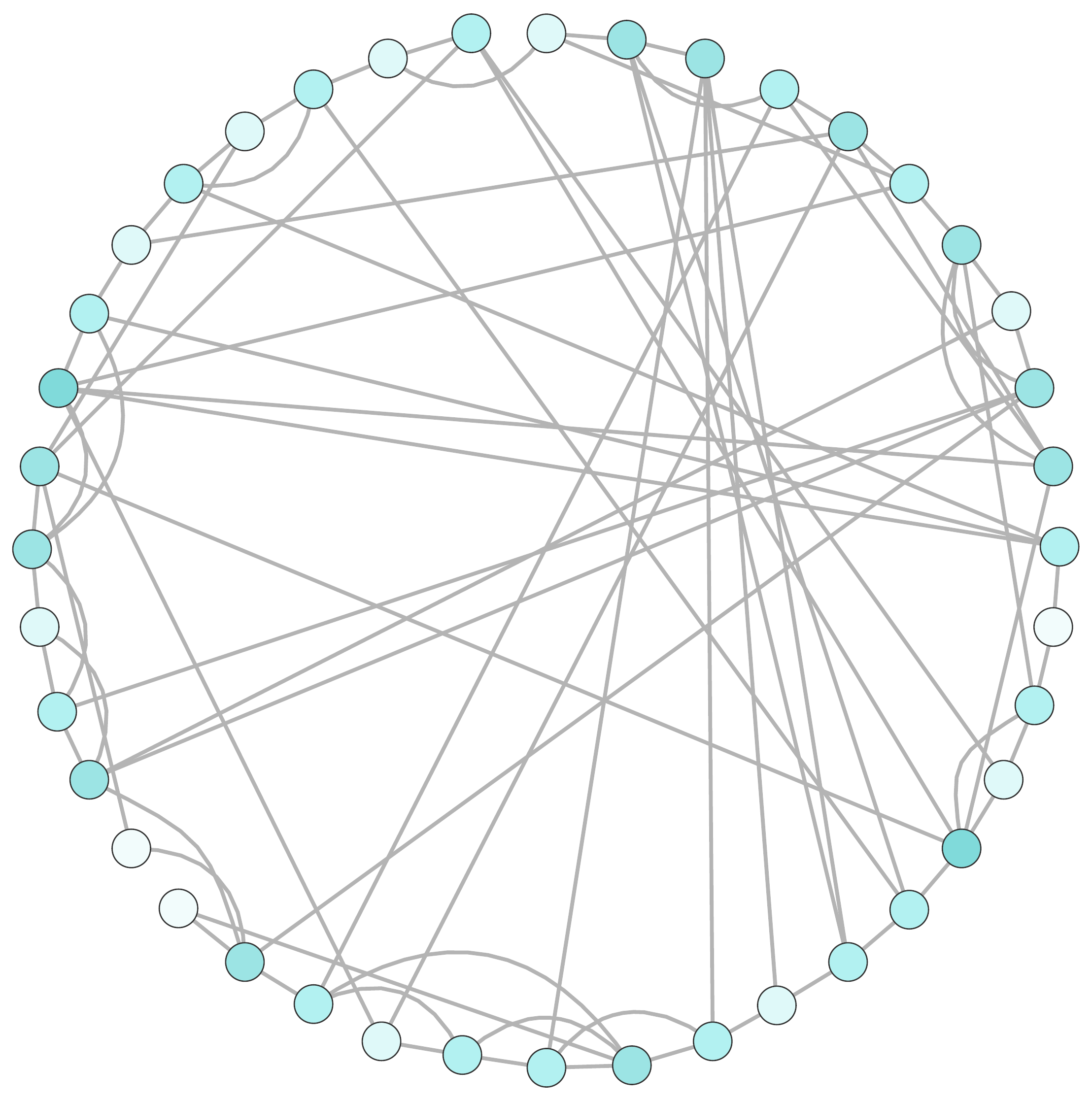}
  \caption{Watts-Strogatz (small world) graph, in which each node connects to its 4 nearest neighbours. 30\% of edges are then randomised.}
  \label{smallworld}
\end{subfigure}%
\hspace*{.05\textwidth}
\begin{subfigure}{.45\textwidth}
  \centering
  \includegraphics[width=.9\linewidth]{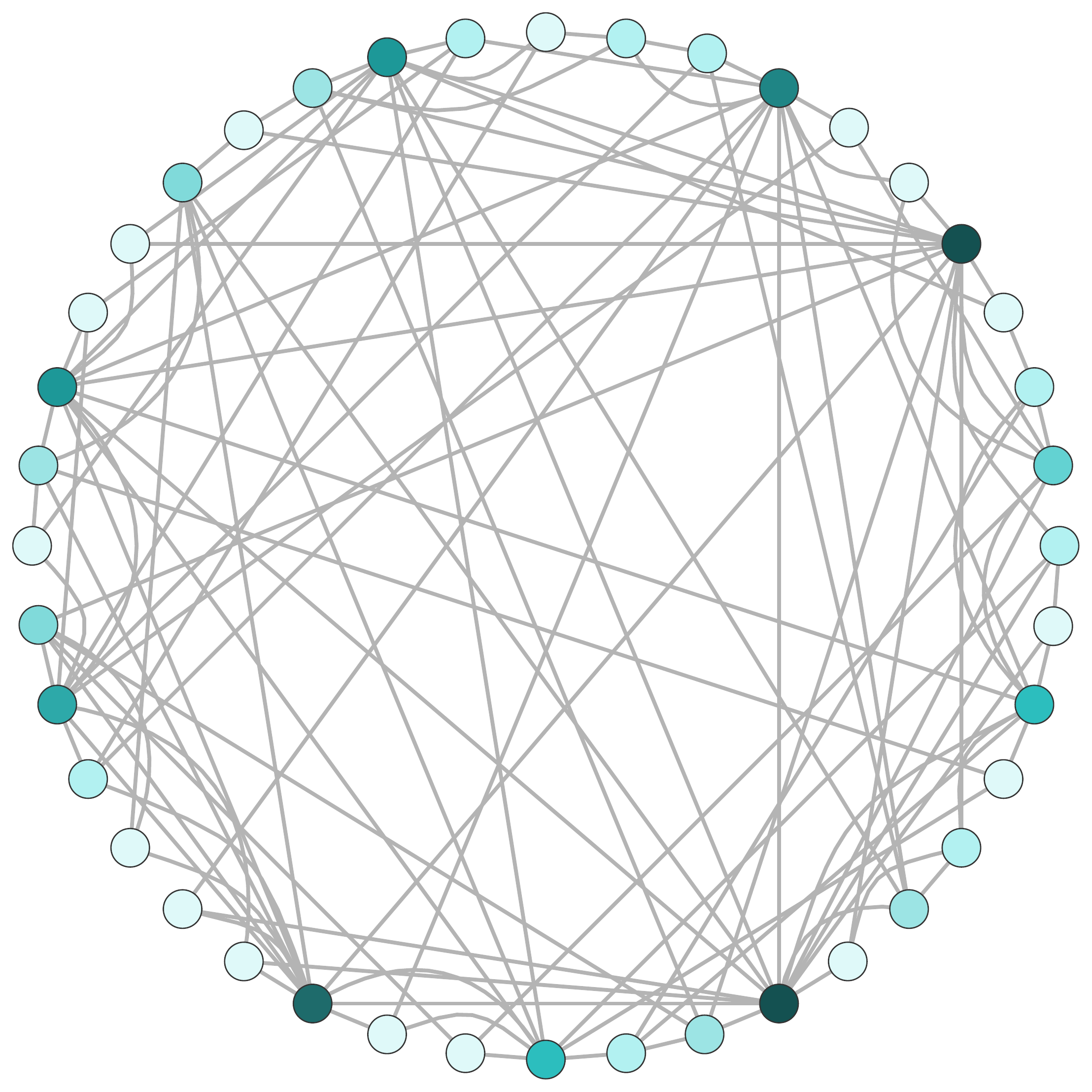}
  \caption{Barab\'asi-Albert graph. A new node connects to three older nodes in the network preferentially based on their degree.}
  \label{BA}
\end{subfigure}
\caption{Comparison between different network models. The intensity of the colour reflects the number of edges adjacent to the node.}\label{farchitecture}
\end{figure}

\subsection{Erd\"os-R\'enyi graph}
The Erd\"os-R\'enyi network model~\cite{ER60} is one of the simplest network models, the essential element of which is its randomness. In this model one assumes no prior knowledge about the network, nodes and edges. To build a representative network of this model, one follows a simple rule: consider each pair of nodes, and with some probability $p$ build an edge between them. 

Erd\"os-R\'enyi network is a ``structure-less'' network, being rather dissimilar from most real-world networks. As such, it typically serves as null model to contrast various properties of the latter. For example, the degree distribution of the real-world networks is often ``fat-tailed" (see \secref{smallworld} for a discussion of networks with fat-tailed degree distribution), which is not the case in an in Erd\"os-R\'enyi graph. Furthermore, the clustering coefficient, as well as typical shortest path lengths tend to be smaller in Erd\"os-R\'enyi than in real-world networks.

\subsection{Lattice graphs}
Lattice graphs are a polar opposite of the Erd\"os-R\'enyi graphs, having an extremely regular, ``symmetric", tile-like shape. To build a lattice graph, one can take a single connection pattern (e.g.\ a triangle, or a square) and repeat it, stitching each element to others in the same way each time.

These networks tend to have very long average path lengths, high clustering coefficient and only one type of dominant motif---the one connection pattern the network is based on. These graphs provide a contrast to the Erd\"os-R\'enyi graph as a null model, as together they are extreme ends of a spectrum within which many real-world networks lie.

\subsection{Small world networks}
Often, the networks which we observe are called ``small-world". The characteristic property of these networks is that most nodes are not immediate neighbours, but any node can be reached from any other node with only a few hops.

One useful way of visualising a small-world network is by depicting it as a circular graph, in which nodes are connected to two other nodes, nearest to them on the circle. This network is a rather ``large world'', as the shortest path between two nodes on the opposite sides of the circle are quite long. To obtain a small-world network from this circular graph one randomly adds a number of ``shortcuts''---i.e.\ edges connecting random pairs of nodes. The inclusion of these shortcuts significantly decreases the length of the shortest path between most pairs of nodes. A network model which yields such properties is the Watts-Strogatz model~\cite{WS98}.

One of the most interesting examples of a small-world network can be found in the well-known Milgram's experiment~\cite{M67}. In it, Stanley Milgram, psychologist at Yale University, sent out packages to some people across the United States. These people were not the recipients of these packages but rather were asked to pass them along to other people they knew well and who may know the recipients (if they did not know them themselves). It turned out that the average path length that a letter travelled was around six---the six degrees of separation. The conclusion was, that, although the social network (people in the country) is large, everyone is separated by a comparably much smaller number of people---six.

\subsection{Scale-free networks}

A network whose degree distribution follows a \vdef{power law} are called scale-free. That means that the probability that a node is observed with a degree $k$ decreases (exponentially) as the degree increases. The term scale-free means that there is no typical number of edges. As a result, we have many nodes with very small degree (close to one), and very few ``winners''---nodes with very large degrees (we say that the degree \vdef{heterogeneity} is large).

Scale-free networks are abundant. De Solla Price was the first to point out that scale-free behaviour is apparent in citation networks~\cite{P65}. These networks are composed of publications as nodes and citations between papers as edges. Thus de Solla Price pointed out that most papers are poorly cited, whereas a few are very well-cited. Other networks that are thought to be scale-free include protein-protein interaction networks, collaboration networks (such as between actors who made films together), the Internet and the World Wide Web to name but a few. In general, systems that are created by ``rich-get-richer'' phenomena tend to be scale-free.

The most famous mechanism to generate scale-free networks is \vdef{preferential attachment}. In preferential attachment, the more connected a node is, the quicker it acquires more new connections. This phenomenon has long been known in non-networks contexts and has many different names, including the Pareto principle or the 80-20 rule ($19^{\mathrm{th}}$ cent.\ ), the Matthew effect (Matthew's gospel: ``For everyone who has will be given more"), Yule's distribution (1925) emerging from a description of biological taxa and subtaxa, and Simon's model for the frequency of words in text.

One of the most well-known generative model for scale-free networks is the Barab\'asi-Albert model of preferential attachment. The growing network model follows very simple rules. It is initiated with a small initial graph. At each time step, a new node is introduced to a network and connect it to $m$ older nodes. The connection happens preferentially, that is, a newly introduced node prefer to connect to large-degree nodes. By growing the network in this way we obtain the scale-free structure\footnote{Please note that while preferential attachment generates edge distributions that follow power laws, the reciprocal is not necessarily true; there are many ways of obtaining power laws that do not relate to preferential attachment.}.

\section{Discussion, future perspectives, and further reading}\label{sdiscussion}

The previous sections have introduced various types of networks, and discussed aspects of how to study them. 
This brief account of network theory certainly is not exhaustive; our goal was to intrigue the reader about the network approach to complex systems, 
as a thorough review of everything this field has to offer is beyond the scope of this work. 

We invite the curious reader to further explore various aspects of network theory in the following texts. Foremost, there are many popular science books that discuss complex networks, including~\cite{FC10,C19,F18,W04}. For more technical, mathematical overview, please refer to handbooks such as~\cite{N10,CC12,E11,L17}. There are also several more specific review papers the on various topic discussed in this article.  For a review of network visualisation see~\cite{WP07} or also explore internet outlets such as \hyperlink{https://towardsdatascience.com/large-graph-visualization-tools-and-approaches-2b8758a1cd59}{Towards Data Science}. For comparison of community detection algorithms, please consider reviews such as~\cite{F10,EATE019,YAT16}. Detailed discussions of centrality measures can be found in~\cite{LFH10,R19}, and visualised in \hyperlink{https://schochastics.net/sna/periodic.html}{schochastics.net}.

Networks provide one of the best and most efficient tools to develop insights over complex datasets. While traditional statistical methods are effective in discovering relationships between variables, network theory is particularly well-suited to study how these relationships are structured, i.e.\ finding ``patterns within the patterns'' In doing so, network theory provide a route towards assessing synergistic properties that might take place within complex systems of interest (c.f. Section~\ref{sec:synergy}). Additionally, networks seem to embody the spirit of our times, being pervasive in the way many things are understood today. Complex networks have come to replace other metaphors in a number of fields of science, and now represent a fundamental element of our worldview.

Having said that, network theory also have important limitations. First, is important to note that a network generated from data is only as good as the data used to build it. In particular, if the data is extremely noisy or untrustworthy, then a network analysis will follow the \textit{garbage-in/garbage-out} principle and provide similarly untrustworthy results. Another important limitation of traditional networks is that they focus on pairwise (i.e. binary) relationships, and then perform inferences on the collective structure of a system based on the relative arrangement of these. However, it has been shown that there exist numerous high-order relationships that cannot be captured by binary measures. These networks can be captured by \textit{hypergraphs}, whose ``hyperlinks'' can relate more than two nodes~\cite{berge1984hypergraphs}. However, the large number of hyperlinks even in relatively small hypergraphs make them computationally challenging. A popular sub-class of hypergraphs, called \textit{simplicial complexes}, have been shown to have favourable properties efficient algorithms for it processing and storage~\cite{edelsbrunner2010computational}. Simplicial complexes are finding diverse applications in recent studies, and we expect them to grow in popularity in the coming years.

Finally, complexity science has many other methods that are not related to networks. Despite its recent popularity, network theory is definitely not a universal solution for social systems and other approaches are sometimes preferable. An overview on complementary methods can be found in general books about complexity science, including~\cite{baryam1999dynamics,thurner2018introduction}.

It is our hope that this review might serve as inspiration of creative uses of network theory and complexity science in the future, fostering novel applications of these tools in fields that have not yet been explored. In particular, we believe that there are plenty of opportunities for using network tools in the fields of arts, both for fostering creative endeavours and guiding analyses of existent material. Nowadays, complexity science is pushing forward the boundaries of our knowledge, and we strongly believe that including arts in this endeavour will bring mutual---and synergistic---benefits to arts and science.

\newpage

\end{document}